\documentclass[%
 reprint,
 amsmath,amssymb,
 aps,
]{revtex4-1}

\usepackage{setspace}
\usepackage{graphicx}
\usepackage{latexsym}
\usepackage{color}
\usepackage[table]{xcolor}
\usepackage{amsmath,amsthm,amssymb}
\usepackage{accents}
\usepackage{flafter}
\usepackage{amssymb}
\usepackage{bbold} 
\usepackage{bm}
\usepackage{pstricks}
\usepackage{pst-plot}
\usepackage{cancel}
\usepackage{eufrak,mathrsfs,mathrsfs}
\usepackage{dsfont}    
\usepackage{makeidx}         
\usepackage{nomencl}         
\usepackage{float}
\usepackage{xfrac}
\usepackage{hyperref}
\usepackage{xcolor}
\definecolor{red}{rgb}{1.0,0.0,0.0}
\definecolor{blue}{rgb}{0.0,0.0,1.0}
\definecolor{dark-gree}{rgb}{0.0,0.5,0.0}
\hypersetup{
    colorlinks, 
    linkcolor={red},
    citecolor={blue}, 
    urlcolor={dark-gree}
}

\usepackage{graphicx}
\usepackage{dcolumn}
\usepackage{bm}

\newcommand{\eq}[1]{\begin{equation}#1\end{equation}}
\newcommand{\tn}[1]{\textnormal{#1}}
\newcommand{\lb}[1]{\label{#1}}

\newcommand{\bra}{\langle}
\newcommand{\ket}{\rangle}
\newcommand{\dst}{\displaystyle}

\begin{document}

\preprint{APS/123-QED}

\title{Quantum speed limit for a relativistic electron in a uniform magnetic field}

\author{D. V. Villamizar}
 \email{david.velasco.v@gmail.com}
\author{E. I. Duzzioni}
 \email{duzzioni@gmail.com}
\affiliation{ Departamento de F\'isica, Universidade Federal de Santa Catarina, Santa Catarina, CEP 88040-900, Brazil}%




\date{\today}

\begin{abstract}
We analyze the influence of relativistic effects on the minimum evolution time between two orthogonal states of a quantum system. Defining the initial state as an homogeneous superposition between two Hamiltonian eigenstates of an electron in a uniform magnetic field, we obtain a relation between the minimum evolution time and the displacement of the mean radial position of the electron wavepacket. The quantum speed limit time is calculated for an electron dynamics described by Dirac and Schroedinger-Pauli equations considering different parameters, such as the strength of magnetic field and the linear momentum of the electron in the axial direction. We highlight that when the electron undergoes a region with extremely strong magnetic field the relativistic and non-relativistic dynamics differ substantially, so that the description given by Schroedinger-Pauli equation enables the electron traveling faster than $c$, which is prohibited by Einstein's theory of relativity. This approach allows a connection between the abstract Hilbert space and the space-time coordinates, besides the identification of the most appropriate quantum dynamics used to describe the electron motion. 
\begin{description}
\item[PACS numbers]
\end{description}
\end{abstract}

\pacs{Valid PACS appear here}
\maketitle

 \vspace{-0.3cm}
\section{\label{sec:introduction}Introduction}
 \vspace{-0.3cm}
The question  \emph{How fast can quantum information be processed?} was tackled at first time in 1945 by Mandelstam and Tamm \cite{MT}. They developed a criterion to find the minimum time for a closed quantum system with limited energy uncertainty $\Delta H$ to change the expectation value of a given operator by the standard of this operator. Such result was supported later by \cite{Fleming, Anandan, Vaidman}. On the other hand, Margolus and Levitin \cite{ML} attributed the speed of a quantum evolution between two orthogonal states to the mean energy of the system $\bra\hat{H}\ket$. In Ref. \cite{Giovannetti_Lloyd} it is assumed that the minimum evolution time has the following expression $T_{\tn{min}}\!=\!\max\{\pi\hbar/2\Delta H,\pi\hbar/2(\bra \hat{H}\ket\!-\!E_{0})\}$, where $E_{0}$  is the lowest energy of one of the states of the superposition. A unified version of the MT and ML bounds were presented in \cite{Levitin}. Recent developments on this subject extended these ideas to include initial mixed states and open quantum system dynamics, obtaining realistic bounds for the speed of quantum processes \cite{Uhlmann,Davidovich,Huelga_2013,Sebastian_Lutz,Zhang, Marvian, Mondal, Pires}.

 The answer to the former question is very important for many areas of quantum physics, including quantum information and computation \cite{Bekenstein_1,Lloyd_1}, quantum metrology \cite{Lloyd_2}, optimal control theory \cite{Giovannetti}, and quantum thermodynamics \cite{Deffner_2010}. 

Although the achievement of an exact expression for the quantum speed limit is of fundamental importance to attain precisely the minimum time of a quantum process, the correct description of the dynamics of the system of interest is as important as the former. Regarding this point we observe that for an accurate description of a system dynamics it is necessary to take into account relativistic effects. In the case of spin $1/2$ particles such as electrons, the Dirac equation is able to accommodate very well quantum mechanics and special relativity. It reproduces accurately the spectrum of the hydrogen atom, provides a natural description of the electron spin, and the existence of antimatter \cite{P_Strange}. The correction to the energy of atomic levels due to fine structure is a beauty example of relativistic effects in low energy quantum systems. Such correction is very small, about five orders of magnitude smaller than the energy values predicted by the non-relativistic Schr\"{o}dinger equation, nevertheless, experimentally it is observable \cite{Pekka}.

The target of this work is to encompass relativistic effects on the quantum speed limit. For this purpose we analyze the transition between two orthogonal states of an electron in a uniform magnetic field according to Dirac equation \cite{P_Strange} and compare it to the non-relativistic description given by Schroedinger-Pauli equation \cite{Tannoudji,Yoshioka}. Defining the electron initial state as an homogeneous superposition of two eigenstates of the Hamiltonian, the Madelstam-Tamm and Margolus-Levitin bounds become equivalent \cite{Jonas}. Therefore, in some sense, our results are independent on the expression used to calculate the minimum transition time. For some states the electron mean radial position is initially different from its final one. The ratio between such average radial displacement and the minimum evolution time furnishes the average speed in which the electron travels in space-time in the radial direction. Such speed is important for two reasons: \emph{i)} it enables to find what kind of initial superposition state provides the greatest spatial displacement in the shorter time; and \emph{ii)} for speeds higher than the speed of light in vacuum $c$ it works as a criterion to invalidate the equation used to describe the electron dynamics. As expected, the Schroedinger-Pauli equation will be the only one to violate this criterion.

This paper is organized as follows, in section \ref{sec:model_and_framework} we breafly describe the relativistic and non-relativistic dynamics of an electron in a uniform magnetic field by the Dirac and Schroedinger-Pauli equations, respectively. In section \ref{sec:discussion} we show an analysis of a particular case of an initial superposition state which gives us enough information about both quantum mechanical descriptions and used it in base to realize in section \ref{fastest} a numerical calculation for looking for fastest superposition states. In section \ref{conclusion} follows our conclusion.  
\section{MODEL AND FRAMEWORK} \lb{sec:model_and_framework}
For didactic reasons we briefly review the non-relativistic and relativistic dynamics of an electron in a uniform magnetic field, respectively. The Pauli Hamiltonian is
  \eq{\lb{eq:A1}
H = \frac{1}{2m_{0}}\big( \vec{p} + e\vec{A} \big)^{2} + \frac{e}{m_{0}}\vec{B}\!\cdot\!\vec{S},
}
with $\vec{p}$ being the linear mechanical momentum, $e$ the absolute value of the electron charge, and $m_{0}$ the electron rest mass. The magnetic vector potential $\vec{A}$ is expressed by the symmetric Landau gauge $\vec{A}\!=\!(\vec{B}\times\vec{r})/2$, where $\vec{B}=B\hat{z}$ is the magnetic field oriented in $z$ direction and $\vec{r}$ is the vector position of the electron. The eigenstates of the Hamiltonian (\ref{eq:A1}) are \cite{Tannoudji,Yoshioka}
 \eq{\lb{eq:A4}
 \psi_{n,m_{l},m_{s},p}(\varrho,\varphi,z) = F_{n,m_{l}}(\varrho, \varphi)e^{ipz/\hbar}\Gamma_{m_{s}}, 
} 
where the radial wavefunction is
\eq{\lb{eq:A25}
\begin{split}
 F_{n,m_{l}}&(\varrho,\varphi) = \frac{\dst(-1)^{\left(\frac{n-|m_{l}|}{2}\right)}\left(\dst\frac{n-|m_{l}|}{2}\right)!}{\sqrt{\pi\left(\dst\frac{n+|m_{l}|}{2}\right)!\left(\dst\frac{n-|m_{l}|}{2}\right)!}} \\
	&\times\beta \big(\beta\varrho\big)^{|m_{l}|} {\large L}^{|m_{l}|}_{\left(\frac{n-|m_{l}|}{2}\right)}(\beta^{2}\varrho^{2}) e^{-\beta^{2}\varrho^{2}/2} e^{im_{l}\varphi},
\end{split}
}
with ${\large L}^{|m_{l}|}_{\left(\frac{n-|m_{l}|}{2}\right)}(\beta^{2}\varrho^{2})$ being the generalized Laguerre polynomials, 
 \vspace{-0.3cm}
\eq{\lb{eq:A25a}
 {\large L}^{|m_{l}|}_{\left(\frac{n-|m_{l}|}{2}\right)} \!\!\!=\!\!\!\!\!\! \sum^{\left(\frac{n-|m_{l}|}{2}\right)}_{j=0} \!\!\!\!\!\!(-1)^{j} \!\!\left(\!\!\begin{array}{c}\left(\frac{n-|m_{l}|}{2}\right)+|m_{l}|\\ \left(\frac{n-|m_{l}|}{2}\right)-j\end{array}\!\!\right) \frac{\dst\big(\beta\varrho\big)^{2j}}{j!}.
}
Here $\beta \equiv \sqrt{\frac{eB}{2\hbar}}$ is the inverse of the \textit{characteristic length} of the harmonic oscillator, the indexes $n=0,1,2,...$ and $m_{l}=-n,-n+2,...,n-2,n$ refer to the eigenstates $F_{n,m_{l}}(\varrho, \varphi)$ of the 2-dimensional harmonic oscillator in the plane perpendicular to the orientation of the magnetic field and also to the coupling between the magnetic field and the orbital angular momentum. $\Gamma_{m_{s}}$ represents the eigenstates of the spin opetator $S_{z}$ with eigenvalues $\hbar m_{s}$, so that the index $m_{s}=\{-1/2,+1/2\}$. $p$ is the projection of the linear momentum in $z$ direction. The corresponding eigenvalues of Hamiltonian (\ref{eq:A1}) are 
\eq{\lb{eq:A27}
 E_{n,m_{l},m_{s},p} = \frac{p^{2}}{2m_{0}} + \hbar\omega\big( n + m_{l} + 2m_{s} + 1 \big), \hspace{0.5cm} \omega \equiv \frac{eB}{2m_{0}}.
} 

By its turn, the relativistic dynamics of the electron is given by Dirac equation, which one is expressed as 
\eq{\lb{eq:d-1}
 i\hbar\dfrac{\partial}{\partial t}\psi(\vec{r},t) \!=\! \big( c \vec{\alpha}\!\cdot\!\vec{\Pi} + \beta m_{0}c^{2} \big)\psi(\vec{r},t),
}
where $\vec{\Pi} \!=\! \vec{p} + e\vec{A}$ is the linear canonical momentum. We are using the Bjorken-Drell convention to represent the $\gamma$ matrices, here denoted by $\vec{\alpha}$ and $\beta$. The Dirac Hamiltonian eigenstates are spinors with four components, in which the two upper components have positive energy and are described by Eq.(\ref{eq:A4}), while the two lower components with negative energy are given by 
\eq{
\frac{c \vec{\sigma}\!\cdot\!\vec{\Pi}}{E+m_{0}c^2}\psi_{n,m_{l},m_{s},p}(\varrho,\varphi,z).
}
The quantity $E$ represents the eigenenergies 
\eq{\lb{eq:d-14}
 E_{n,m_{l},m_{s},j,p} \!=\! j\sqrt{ m^{2}_{0}c^{4} \!+\! p^{2}c^{2} \!+\! eB\hbar c^{2}\big( n \!+\! m_{l} \!+\! 2m_{s} \!+\! 1 \big) },
}
with  $j=\{+,-\}$ indicating the sign of the energy. For more details about this solution see Ref. \cite{P_Strange}.

The electron initial state is assumed to have $+1/2$ spin projection along the $z$ direction and a gaussian wave packet in the same spatial direction with standard deviation $d$ and expectation value $p_{0}$ for the linear momentum operator $\hat{p}_{z}$, 
\begin{equation}\lb{eq:gauss}
 \psi_{z}(z) = \frac{1}{\big(2\pi d^{2}\big)^{1/4}} e^{-z^{2}/4d^{2}} e^{ip_{0}z/\hbar}.
\end{equation}

Our idea is to establish a connection between the quantum speed limit and the speed in which the electron wave packet moves through the space-time. For this purpose we consider the initial state of the system in $x\!-\!y$ plane as a homogeneous superposition of two radial eigenstates $F_{n,m_{l}}(\varrho,\varphi)$ in different Landau energy levels. After the time of evolution $T_{\tn{min}}$ the state of the system is orthogonal to the initial one, so that the mean radial position of the electron wave packet experiences a displacement. In the next sections we analyze the relativistic effects on $T_{\tn{min}}$, besides the dependence of the electron's displacement on the initial superposition state and on the relativistic and non-relativistic descriptions of quantum mechanics.
\section{Quantum speed limit for an electron under relativistic and non-relativistic quantum dynamics} \lb{sec:discussion}
We start analyzing the non-relativistic case of superposition between the radial eigenstates $F_{0,0}(\varrho,\varphi)$ and $F_{2,0}(\varrho,\varphi)$, 
\eq{\lb{eq:D1}
 \psi(\vec{r},0) = \frac{1}{\sqrt{2}} \left[ F_{0,0}(\varrho,\varphi) + F_{2,0}(\varrho,\varphi) \right] \psi_{z}(z)\Gamma_{+1/2}.
}
After the evolution from $\psi(\vec{r},0)$ to $\psi(\vec{r},T_{\tn{min}})$, we obtain the quantities required to evaluate the quantum speed limit criteria,
\eq{\lb{eq:D4}
 \Delta H \!=\!  \bra H\ket \!-\! E_{0}  \!=\! \frac{eB\hbar}{2m_{0}}.
}
Thus the minimum evolution time is,
\eq{\lb{eq:D5}
 T_{\tn{min}} = \frac{\pi m_{0}}{eB}.
}
As we are considering basically the change in the radial part of the system state, we will analyze the radial displacement of the electron, which enable us to set the expectation value of the linear momentum in the axial direction as $p_{0}=0$. Thus, the mean radial position at any time is given by the expression,
\eq{\lb{eq:D6}
\begin{split}
 \bra\varrho\ket_{t} &\!=\! \frac{1}{2} \Big[ \bra0,0|\varrho|0,0\ket \!+\! \bra2,0|\varrho|2,0\ket \\
 	 &\hspace{2.4cm} 	+\! 2\bra0,0|\varrho|2,0\ket \cos\big(\mathcal{E}t\big)  \Big].\\
\end{split}
}
By using the Dirac notation, we nominated each eigenstate by its quantum numbers $n$ and $m_{l}$, and $\mathcal{E}\!=\!eB/m_{0}$ is a constant with dimension of frequency. Therefore, the maximum radial displacement of the electron's mean position is,
\eq{\lb{eq:D7}
\begin{split}
 \big| \bra\varrho\ket_{T_{\tn{min}}} \!-\! \bra\varrho\ket_{0} \big| &= \Big| \bra0,0|\varrho|2,0\ket \big[\cos(\mathcal{E}T_{\tn{min}}) \!-\! 1\big] \Big|. \\
	&= \sqrt{\frac{\pi\hbar}{2eB}}.
\end{split}
}
The expression above shows us the relevancy of the crossed term $D_{S}(\varrho) \!=\! \bra0,0|\varrho|2,0\ket \!=\! 2\pi \int_{0}^{\infty} F_{0,0}^{\dagger}(\varrho,\varphi) \varrho F_{2,0}(\varrho,\varphi) \varrho d\varrho$  for the electron's displacement. From Eqs. (\ref{eq:A4}), (\ref{eq:A25}), and (\ref{eq:D7}) we observe that for a non null radial displacement of the electron, the initial superposition state must be built by eigenstates with the same quantum numbers of spin $m_{s}$ and orbital angular momentum $m_{l}$, besides $\mathcal{E}T_{\tn{min}} \neq s\pi$, with $s$ even. Then the average speed of the mean radial position of the electron from its initial state to the orthogonal one is,
\eq{\lb{eq:D8}
 \bar{v} = \frac{1}{m_{0}}\sqrt{\frac{eB\hbar}{2\pi}}.
}\\
On the other hand, in the relativistic description with $p_{0}=0$, the minimum evolution time is
\eq{\lb{eq:D10}
 T_{\tn{min}} \!=\! \frac{\pi\hbar }{\sqrt{m^{2}_{0}c^{4} \!+\! 4eB\hbar c^{2}} \!-\! \sqrt{m^{2}_{0}c^{4} \!+\! 2eB\hbar c^{2}}}.
}
At follows we write the two spinors that compose the evolved state of the system 
\eq{\lb{eq:D13}
 U_{0,0} = N_{0,0} \left(\begin{array}{c}  F_{0,0} (\varrho,\varphi) \\ 0 \\ \dst\frac{cpF_{0,0}(\varrho,\varphi)}{(E_{0,0}\!+\!m_{0}c^{2})} \\ \dst\frac{2i \hbar c\beta F_{1,1}(\varrho,\varphi)}{(E_{0,0} \!+\! m_{0}c^{2})} \end{array}\right)e^{ipz/\hbar},
}
and 
\eq{\lb{eq:D14}
 U_{2,0} \!=\! N_{2,0}\! \left(\!\begin{array}{c} F_{2,0}(\varrho,\varphi) \\ 0 \\ \dst\frac{cpF_{2,0}(\varrho,\varphi)}{(E_{2,0}\!+\!m_{0}c^{2})} \\ \dst\frac{2i \hbar c\beta\sqrt{2}F_{3,1}(\varrho,\varphi)}{(E_{2,0}\!+\!m_{0}c^{2})} \end{array}\right)\! e^{ipz/\hbar},
}
where $N_{0,0}$ and $N_{2,0}$ are normalization constants, and $E_{0,0}$ e $E_{2,0}$ are positive eigenvalues given by Eq. (\ref{eq:d-14}). Therefore, the superposition state evolves in time as,
\eq{\lb{eq:D15}
 \psi(\vec{r},t) \!\!=\!\!\! \frac{1}{\sqrt{2}} \int_{-\infty}^{\infty} \alpha(p) \Big[\! U_{0,0}e^{-iE_{0,0}t/\hbar} \!+\! U_{2,0}e^{-iE_{2,0}t/\hbar}  \!\Big]dp,
}
with $\alpha(p)$ being the coefficient of expansion of the Gaussian wave packet defined in Eq.(\ref{eq:gauss}). Now we are able to calculate the radial displacement of the electron's mean position in the relativistic case, which one is made numerically \cite{Fortran}. In FIG. (1) we plot the average speed of the electron's wave packet when moving from the initial to final state during the time interval $T_{\tn{min}}$ under both relativistic and non-relativistic quantum dynamics. We noticed in the non-relativistic case that there is a magnetic field strong enough to yield $\bar{v} \geq c$ given by $B \geq 2.77\times\!10^{10}$T, which contradicts the Einstein's theory of relativity. 
\begin{figure}[h]
 \centering
  \includegraphics[scale=0.38]{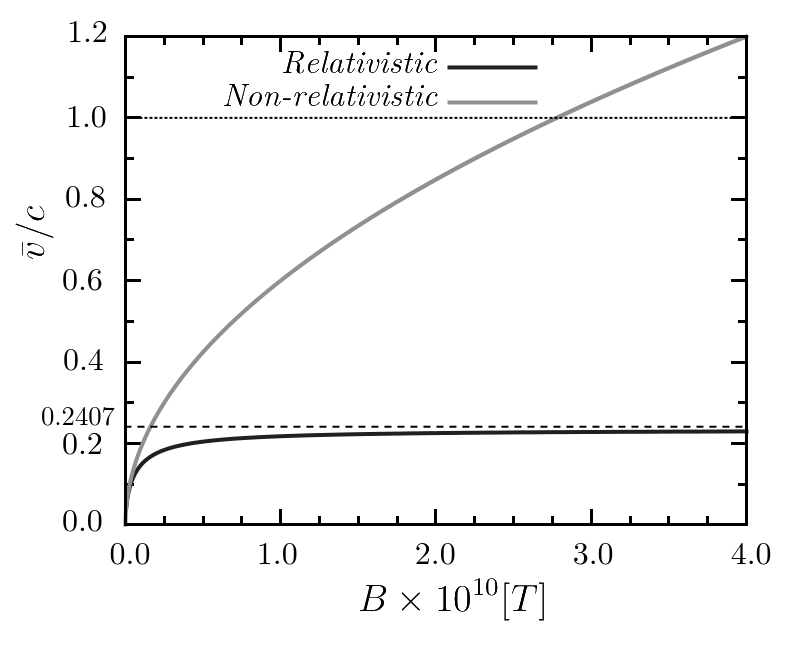}
 \label{fig:both}
 \vspace{-0.3cm}
 \caption{\small Average radial speed as function of the external magnetic field according to both quantum dynamics.}
\end{figure}
Naturally, it is impossible to achieve this intensity of magnetic field in a laboratory on the Earth, but not in special neutron stars, called magnetar \cite{Harding}. Conversely, the Dirac's theory for the electron predicts the asymptotic value of $\bar{v} \backsimeq 0.2407c$. To attain this value we first need to evaluate the radial displacement of the electron mean position, which one depends on the crossed term $D_{S}(\varrho) \!=\! 2\pi U^{\dagger}_{0,0}\varrho U_{2,0}$ and on the minimum evolution time $T_{\tn{min}}$ as in Eq. (\ref{eq:D10}) . In the limit case $B\!\rightarrow\!\infty$ the expressions for the eigenenergies can be approximated by
\eq{\lb{eq:D17}
  E_{0,0} \approx 2c\hbar\beta, \quad    E_{2,0} \approx 2\sqrt{2}c\hbar\beta.
}
which renders,
\eq{\lb{eq:D18}
 T_{\tn{min}} \approx \frac{\pi}{2 c \beta(\sqrt{2}-1)}.
}
Inside this approximation, the spinor normalization constants become $N_{0,0}\!=\!N_{2,0}\!=\!1/\sqrt{2}$ and the radial displacement of the electron's mean position becomes,

\vspace{-0.5cm}
\eq{\lb{eq:D20}
\begin{split}
 \big| \bra\varrho\ket_{T_{\tn{min}}} \!-\! \bra\varrho\ket_{0} \big| &= 2\Bigg|\int\limits^{\infty}_{0}\varrho D_{S}(\varrho)d\varrho\Bigg|, \\
	& = \frac{\sqrt{\pi} }{4\beta} \left( 1 \!+\! \frac{3}{2\sqrt{2}} \right).
\end{split}
}
Differently from the non-relativistic case, now the displacement in time and space have the same dependence on the magnetic field, as shown in Eqs. (\ref{eq:D18}) and (\ref{eq:D20}), respectively. Consequently, the maximum value of the average speed of the electron in the radial direction is
\eq{\lb{eq:D21}
 \bar{v} = \frac{c}{4\sqrt{2\pi}}\big(1+\sqrt{2}\big) \approx 0.2407c.
}
Hence, as expected, the relativistic quantum dynamics is the most appropriate to describe the electron dynamics in the presence of high intensity magnetic fields. Throughout the present development we observed that the relativistic theory of quantum mechanics funded by Dirac does not restrict the time interval of a quantum process of being arbitrarily small, as shown in Eq. (\ref{eq:D10}). Making a comparison between Eqs. (\ref{eq:D5}) and (\ref{eq:D10}) we verify that there is a quadratic dependence on the magnetic field in the non-relativistic case in relation to the relativistic one. Such difference can turn out to be important for $B \sim m_{0}^{2}c^{2}/e\hbar \sim 5$GT. However, our relativistic description of the electron dynamics, and consequently the quantum speed limit, applies for low intensity magnetic fields in graphene, where the charge carriers can effectively be described by relativistic particles with zero rest mass \cite{Novoselov, Peres_2006}.  
 \vspace{-0.3cm}
\section{The fastest superpositions}\lb{fastest}
In the preceding section we verified that the dynamics of an electron described by Schroedinger-Pauli equation violates a basic principle of Einstein's theory of relativity, which states that any object with non-null rest mass cannot travel faster than $c$. For that reason, we study the dependence of the quantum speed limit for an electron evolving according to the Dirac theory as function of the initial superposition state. Our main purpose here is looking for the maximum radial displacement in the shortest time interval. Since the electron's radial displacement depends strongly on the crossed term $D_{S}(\varrho)$, its maximum absolute value is attained when the initial and final states have the same spin orientation ($m_{s}\!=\!1/2$), zero angular momentum projection $m_{l}\!=\!0$, and the initial superposition state is made of two nearest neighbors eigenstates, i.e., with quantum numbers $n$ and $n+2$. In FIG. 1 we observe that $\bar{v}$ increases as the intensity of the magnetic field is strengthened. Therefore, in the regime $\beta\!\rightarrow\!\infty$ the minimum evolution time between two orthogonal states, where the initial superposition state is composed by two eigenstates with positive energy (called \textit{particle-particle} states), is
\eq{\lb{eq:ext_2}
 T_{\tn{min}} \approx \frac{\pi}{\big[\sqrt{n\!+\!4} \!-\! \sqrt{n+2} \big]\sqrt{2} c\beta},
}
and the crossed term turns out to be, 
\eq{\lb{eq:ext_3}
 D_{S}(\varrho) \approx \pi\varrho \Big[ F^{\dagger}_{n,0}F_{n+2,0} + F^{\dagger}_{n+1,1}F_{n+3,1} \Big].
}
After some steps we get an analytic expression for the maximum radial displacement as function of $n$
\begin{widetext}
\eq{\lb{eq:ext_4}
\big| \bra\varrho\ket_{T_{\tn{min}}} \!-\! \bra\varrho\ket_{0}  \big| \!=\! \frac{1}{\beta}\!\sum^{\left(\frac{n}{2}\right)}_{i=0} \!\sum^{\left(\frac{n+2}{2}\right)}_{j=0} \!\frac{(-1)^{i+j}}{i! \;j!} \!\left(\!\begin{array}{c} \frac{n}{2} \\ \frac{n}{2} \!-\! i \end{array}\!\right)  \!\left(\!\begin{array}{c} \frac{n+2}{2} \\ \frac{n+2}{2} \!-\! j \end{array}\!\right)\! \Gamma\!\!\left(\!i\!+\!j\!+\!1\!+\!\frac{1}{2}\!\right) \left[ 1 \!+\! \frac{\sqrt{ \left(\frac{n}{2}\!+\!2\right)\left(\frac{n}{2}\!+\!1\right)}}{(i\!+\!1)(j\!+\!1)} \! \left(\!i\!+\!j\!+\!1\!+\!\frac{1}{2}\!\right)\! \right].
}
\end{widetext} 
In FIG.2 we plot the average radial speed of the electron for different even values of $n$ ranging in the interval $[0,132]$. The evaluation of $\bar{v}$ for higher values of $n$ is very hard, provided that Eq. (\ref{eq:ext_4}) has many factorials. The inset of such figure shows the convergence of $\bar{v}/c$ to the asymptotic value $0.269814$ found numerically. We notice that in the interval $80 \!\leq \!n\! \leq\! 132$  the value of $\bar{v}$ changes in the fourth decimal place only, which shows that the average radial speed is reaching a constant value less than $c$ for $n\!\rightarrow\!\infty$. 

\begin{figure}[h!]
 \centering
  \includegraphics[scale=0.27]{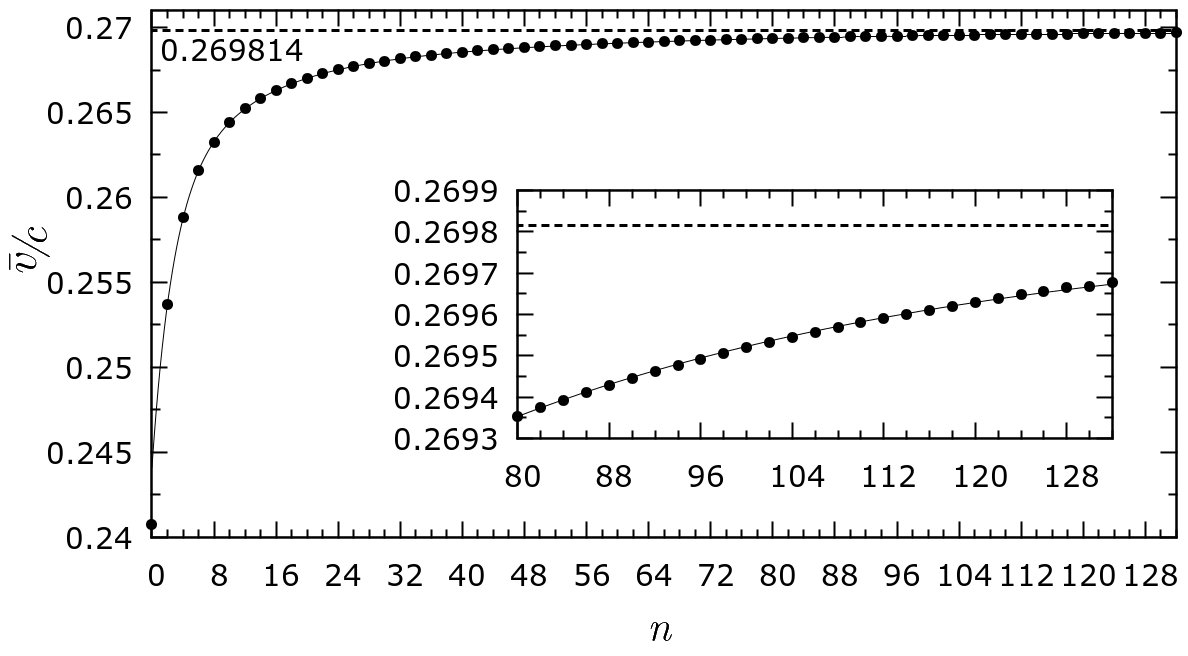}
 \lb{fig:asymptote_positive}
 \vspace{-0.4cm}
 \caption{\small Average radial speed of an electron for different initial superpositions of two positive energy eigenstates $\left( U_{n,0} + U_{n+2,0} \right)/\sqrt{2}$.}
\end{figure}

Instead of considering only initial particle-particle states, we will take into account superpositions of eigenstates with negative and positive energies (called \textit{antiparticle-particle} states). The reason we are tackling this subject only now is that it is not clear if it is fair to compare the non-relativistic dynamics, which describes only particle states, with the relativistic antiparticle dynamics. Despite that, antiparticle-particle dynamics reveals the role played by the electron rest mass in the energy spectrum and thus imposes physical limits on the quantum speed limit \cite{Lloyd}. Repeating the same procedure above to obtain the maximum displacement of the mean radial position of the electron, we find  that the two states of the superposition must have the same spin orientation (spin up), null angular momentum projection along the $z$ direction, and must be made of nearest neighbors eigenstates with even quantum numbers $n$. Assuming the negative energy eigenvalue as the lowest one in module, according to Eq. (\ref{eq:d-14}) we attribute to it the quantum number $n$, while for the positive energy eigenvalue the quantum number $n+2$. Thus, the minimum evolution time for the particular case $n=0$ and $p_{0}=0$ is given by
\begin{equation}\lb{eq:mim-neg}
T_{\tn{min}} \!=\! \frac{\pi\hbar }{\sqrt{m^{2}_{0}c^{4} \!+\! 4eB\hbar c^{2}} \!+\! \sqrt{m^{2}_{0}c^{4} \!+\! 2eB\hbar c^{2}}}.
\end{equation}
This time is shorter than in the particle-particle case (see Eq. (\ref{eq:D10})) because the energy gap is bigger by a quantity that is at least the energy of the electron rest mass. To evaluate the mean radial displacement of the electron, we need the expression of the negative energy spinor for a general quantum number $n$ and null angular momentum, 
\eq{\lb{eq:np2}
 U_{n,0}^{-}(\vec{r}) = N_{n,0}^{-} \left(\begin{array}{c} \dst \frac{cp}{(E_{n}\!-\!m_{0}c^{2})} F_{n,0} \\ \dst \frac{\sqrt{2}ic\hbar\beta\sqrt{n+2}}{(E_{n}\!-\!m_{0}c^{2})} F_{n+1,1}  \\  F_{n,0} \\ 0 \end{array}\right) e^{ipz/\hbar},
}
where $N_{n,0}^{-}$ is the normalization constant. In addition, the radial displacement of the electron is proportional to the absolute value of the crossed term 
\eq{\lb{eq:np3}
\begin{split}
\!\!\!\!  D_{S} \!=\! 2\pi\varrho & N_{n,0}^{-}N_{n+2,0} cp \\
	 &\times\! \bigg[\!\!\frac{1}{E_{n}\!-\!m_{0}c^{2}}+\frac{1}{E_{n+2}\!+\!m_{0}c^{2}}\!\!\bigg] F^{\dagger}_{n,0}F_{n+2,0},
\end{split}
}
which one is maximized for $p_{0}\!\approx\! \beta \hbar \gg m_{0}c$. In FIG. 3 we plot the average radial speed of the electron to change from a $n$ dependent initial negative-positive state to a final one orthogonal to the former in the minimum time interval $T_{\tn{min}}$. The asymptotic value of $\bar{v}$ is $0.134743c$ and lower than the speed in the positive-positive case (see FIG. 2). 
\begin{figure}[h!]   
 \centering
  \includegraphics[scale=0.26]{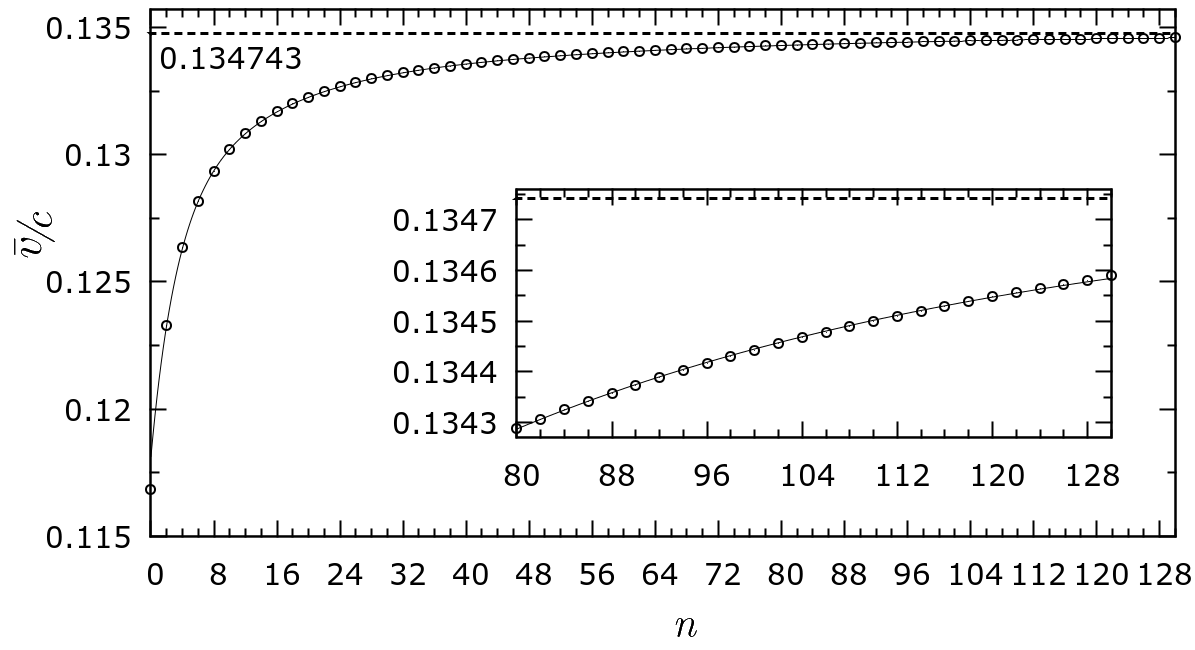}
 \lb{fig:asymptote_negative}
 \vspace{-0.4cm}
 \caption{\small  Average radial speed of an electron for different initial superposition states composed by a positive and a negative energy eigenstate $\left( U_{n,0}^{-} + U_{n+2,0} \right)/\sqrt{2}$.}
\end{figure}
\\
Making a comparison between FIGs. 2 and 3 we observe that $\bar{v}$ for negative-positive states is always less than $\bar{v}$  for positive-positive states. This behavior is clarified in FIG. 4, where $\bar{v}/c$ is plotted for both cases of initial superposition states as function of $p_{0}$ for three different values of the magnetic field.  If the initial state is negative-positive, then, according to Eq. (\ref{eq:np3}), the displacement of the radial mean position of the electron depends linearly on $p_{0}$, which justify the null value of $\bar{v}/c$  at the origin of FIG. 4. For intermediate values of $p_{0}$, we observe that $\bar{v}/c$ attains a maximum value for $p_{0}\!\approx\! \beta \hbar \gg m_{0}c$, while for great values of $p_{0} \gg \beta \hbar, m_{0}c$ whatever the initial superposition the average radial speed becomes smaller. In the later case both positive and negative energy eigenstates have the same expression, and therefore the same radial displacement of the electron and $T_{\tn{min}}$. 

In the context of Dirac's theory, this can be explained by the fact that each spinor does not describe its own particle only, but also its antiparticle by the two terms in the bottom position of the spinor. One of these antiparticle terms is relevant to the whole description when the electron presents a very high linear momentum or when the particle is strongly confined in a region less than or equal to its Compton wavelength. In these cases we could say that the spinor by itself describes a superposition between its particle and its antiparticle \cite{Unanyan_2009,Cheng_2011,P_Strange}.
\begin{figure}[h!]
 \centering
  \includegraphics[scale=0.36]{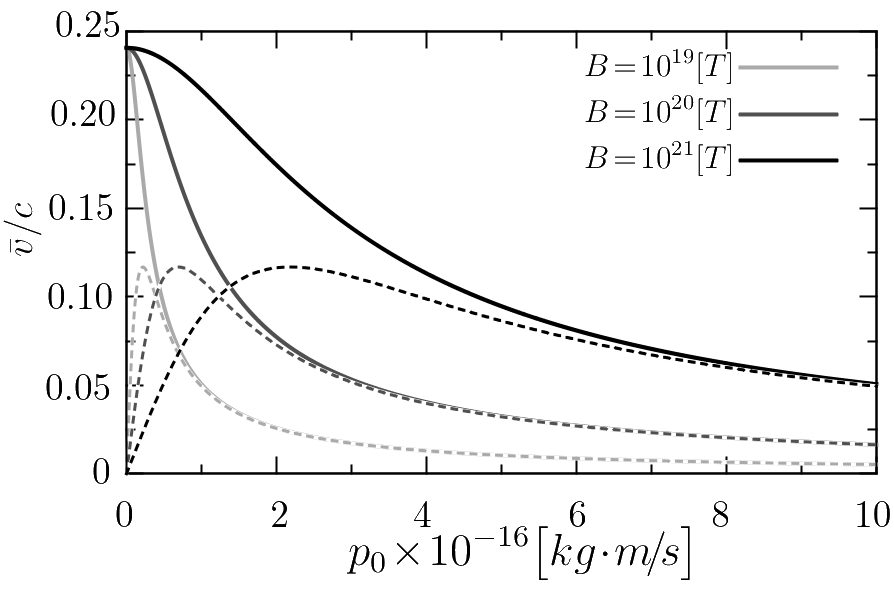}
 \lb{fig:positivo_negativo}
 \vspace{-0.5cm}
 \caption{\small Average radial speed of an electron for particle-particle (solid lines) and antiparticle-particle (dashed lines) states as function of the expectation value of the linear momentum along the $z$ direction, $p_{0}$.}
\end{figure}
\section{CONCLUSIONS}\lb{conclusion}

We analyzed the role played by relativistic effects on the quantum speed limit of a system composed by an electron in a uniform magnetic field. The relativistic dynamics by itself does not restrict the minimum time of evolution of being arbitrary small, but imposes constraints on the average speed at which the electron travels along the space-time. As expected, we observed that the quantum dynamics described by Schroedinger-Pauli equation enables the electron wave packet traveling faster than $c$, in contradiction to Einstein's theory of relativity. Such problem is circumvented by the use of Dirac's equation. The minimum evolution time between two orthogonal states in the relativistic formulation can be significantly different from the non-relativistic case. If the initial state is a homogenous superposition of two Hamiltonian eigenstates with positive energies, then, the minimum evolution time is dilated in the laboratory frame. On the other hand, if the Hamiltonian eigenstates have negative and positive energies, then, the minimum evolution time is contracted in the laboratory frame. This last result can be useful for quantum computing, since it can speed up quantum gates, although a precise control over the creation of particle-antiparticle states is necessary \cite{FG}.

\section{ACKNOWLEDGEMENTS}
 \vspace{-0.3cm}
This work is part of the Brazilian National Institute for Science and Technology of Quantum Information and was
supported by the Brazilian funding agencies CNPq and CAPES.

\bibliographystyle{unsrt} 

\begin{thebibliography}{99}
\bibitem{Fortran}{All numerical calculations made in this manuscript were performed by using FORTRAN 95.}
\end{thebibliography}
\addcontentsline{toc}{chapter}{\protect\numberline{}Bibliografia}

\end{document}